\documentclass[
 twocolumn,
superscriptaddress,
 amsmath,amssymb,
 aps,
pra,
]{revtex4-1}

\usepackage{graphicx}
\usepackage{dcolumn}
\usepackage{bm}
\usepackage{tikz}
\usepackage[colorlinks=true,urlcolor=blue,citecolor=blue,linkcolor=blue]{hyperref}
\usepackage{orcidlink}
\usepackage{comment}
\usetikzlibrary{quantikz}
\usepackage{physics}
\usepackage{amsfonts}
\usepackage{dsfont}
\usepackage{algorithm}
\usepackage{algorithmic}
\usepackage{braket}
\usepackage{ragged2e}
\usepackage{amsmath}
\usepackage{mathtools}
\usepackage[mathlines]{lineno}
\usepackage[normalem]{ulem}
\usepackage{amsthm}
\usepackage{multirow}
\usepackage{bm}
\usepackage{dsfont}
\usepackage{makecell}
\usepackage{amssymb}
\usepackage{physics}
\usepackage{mathtools}
\usepackage{xcolor}\usepackage{soul}
\usepackage{etoolbox}
\usepackage{graphicx}
\usepackage{mwe}
\usepackage{cancel}
\usepackage[prefix]{nomencl} 
\usepackage{etoolbox}        
\usepackage{amsmath,amssymb} 
\usepackage{siunitx}
\usepackage{makecell}

\DeclareMathOperator{\EE}{\mathbb{E}}

\begin{document}

\title{Breaking concentration barriers for quantum extreme learning on digital quantum processors}

\author{Timoth\'ee Dao \orcidlink{0009-0009-2344-3529}}\email{timothee.dao1@ibm.com}\thanks{These authors contributed equally to this work.}
\affiliation{IBM Research Europe -- Zurich, 8803 R\"{u}schlikon, Switzerland}
\affiliation{Institute for Theoretical Physics, ETH Z\"{u}rich, 8093 Z\"{u}rich, Switzerland}

\author{Ege Yilmaz \orcidlink{0009-0005-9098-350X}}\thanks{These authors contributed equally to this work.}
\affiliation{Department of Mathematics and Computer Science, University of Basel, 4051 Basel, Switzerland}
\affiliation{IBM Research Europe -- Zurich, 8803 R\"{u}schlikon, Switzerland}

\author{Ibrahim Shehzad \orcidlink{0000-0002-0617-7195}}
\affiliation{IBM Quantum, T. J. Watson Research Center, Yorktown Heights, New York 10598, USA}

\author{Christophe Pere}
\affiliation{PINQ2, Sherbrooke, QC, Canada}

\author{Kumar Ghosh \orcidlink{0000-0002-4628-6951}}
\affiliation{E.ON Digital Technology GmbH, Hanover, Germany}

\author{Isabelle Wittmann}
\affiliation{IBM Research Europe -- Zurich, 8803 R\"{u}schlikon, Switzerland}

\author{Thomas Brunschwiler}
\affiliation{IBM Research Europe -- Zurich, 8803 R\"{u}schlikon, Switzerland}

\author{Giorgio Cortiana \orcidlink{0000-0001-8745-5021}}
\affiliation{E.ON Digital Technology GmbH, Hanover, Germany}

\author{Corey O’Meara \orcidlink{0000-0001-7056-7545}}
\affiliation{E.ON Digital Technology GmbH, Hanover, Germany}

\author{Stefan Woerner \orcidlink{0000-0002-5945-4707}}
\affiliation{IBM Research Europe -- Zurich, 8803 R\"{u}schlikon, Switzerland}

\author{Francesco Tacchino \orcidlink{0000-0003-2008-5956}}
\email{fta@zurich.ibm.com}
\affiliation{IBM Research Europe -- Zurich, 8803 R\"{u}schlikon, Switzerland}

\begin{abstract}

Reservoir computing leverages rich, non-linear dynamics to process temporal data. Quantum variants promise enhanced expressivity from high‑dimensional Hilbert spaces, yet their practical applicability is hindered by hardware noise and concentration effects that can erase input–output distinguishability at large system sizes. In this work, we present and experimentally demonstrate a Quantum Extreme Learning Machine (QELM) tailored to state‑of‑the‑art superconducting platforms, employing up to 124 qubits and circuits with more than 5,000 two‑qubit gates on IBM Quantum computers. We introduce a practical multi‑objective hyperparameter tuning strategy that jointly monitors observable variability, capacity, and task performance to identify noise‑robust operating points. In addition, we develop a local eigentask analysis that enables computationally efficient feature selection and effective information retrieval. We report evidence of a regime of optimality that is identifiable at small scales and transferable across tasks and larger systems, and we achieve performances competitive with leading classical baselines on representative benchmarks for time‑series forecasting and satellite image classification. Together, our results establish a viable and robust framework for large‑scale, pre‑fault‑tolerant quantum machine learning and provide a foundation for extending reservoir‑based methods to more expressive architectures and real‑world scenarios.
\end{abstract}

\maketitle    

Reservoir Computing (RC), a framework that processes inputs through the dynamics of a non-linear system~\cite{lukovsevivcius2012reservoir,yan2024emerging}, represents a prominent concept in classical machine learning, particularly for the analysis of sequential data. In recent years, the concept of RC has been extended to quantum systems, with the aim of exploiting their complex dynamics and high-dimensional Hilbert spaces~\cite{fujii2021quantum,mujal2021opportunities,gyurik2026quantum} while, at the same time, trying to avoid some of the practical and theoretical pitfalls of other popular Quantum Machine Learning (QML) methods~\cite{larocca2025barren}. 

Following this growing interest in Quantum Reservoir Computing (QRC), many works have focused on analog implementations~\cite{negoro2021toward,kornjavca2024large,cimini2025large} as well as simulations and proof-of-principle experiments on digital platforms~\cite{fujii2017harnessing,martinez2021dynamical,mujal2023time,martinez2023information,kubota2023temporal,yasuda2023quantum,hu2024overcoming,wudarski2024hybrid,kobayashi2024feedback,franceschetto2025harnessing,settino2025memory}. At the same time, a series of theoretical studies~\cite{xiong2025fundamental,xiong2025role,sannia2025exponential} uncovered the existence of concentration phenomena, akin to barren plateaus in quantum neural networks~\cite{larocca2025barren} and to similar results holding for quantum kernel methods~\cite{thanasilp2024exponential}, which may hinder the scalability of QRC techniques to hundreds of qubits, a key prerequisite for quantum advantage. The search for viable QRC architectures suited for current and future digital quantum processors that are capable of addressing real-world machine learning tasks is therefore a key challenge. Besides the difficulty of experimental realization of large scale QRC systems, a key component to understand is the trade-off between circuit expressivity and so-called scrambling dynamics~\cite{martinez2021dynamical,xiong2025fundamental,xiong2025role,sannia2025exponential}, which can render the outputs practically indistinguishable for different inputs at large system sizes. Only recently has the existing literature begun to address this balance~\cite{xiong2025role,sannia2025exponential}.

In this work, we address both the implementation and the model design aspects by proposing and realizing a Quantum Extreme Learning Machine (QELM)~\cite{ghosh2019quantum,innocenti2023potential,suprano2024,xiong2025fundamental,vetrano2025state,kawanabe2026efficient} architecture, a precursor of full QRC models, tailored to state-of-the-art digital quantum computers. Our design is scalable, noise resilient, and incorporates methods to monitor important metrics such as observable variability, capacity, and performance on prototypical benchmark problems. To achieve this, we introduce practical, multi-objective hyperparameter tuning strategies and a novel local signal-to-noise ratio (eigentask) analysis that together can be used to find optimal operating points, thereby balancing expressivity and robustness. Crucially, we report evidence of universal regimes of optimality offering strong performances transferable across diverse tasks and system sizes.

We demonstrate the effectiveness of our proposed setup by realizing QELM experiments on superconducting quantum processors with up to 124 qubits and 5,084 two-qubit gates. We tackle paradigmatic nonlinear autoregressive moving-average (NARMA) time-series benchmarks and application-oriented classification tasks based on the Statlog Landsat Satellite dataset~\cite{uci_statlog_landsat}, both of which are representative of possible future operational use cases under realistic and practically relevant conditions. In the same spirit, we also report numerical simulations of few‑step‑ahead prediction tasks on energy‑market time‑series data. Overall, these findings advance our understanding in pursuit of scalable QML models, by offering a practical and theoretically grounded QELM framework, as well as a promising path towards large-scale deployment of QRC techniques on modern digital quantum processors.

\section{Model design}

An Extreme Learning Machine (ELM)~\cite{wang2022review} is a randomly initialized single-hidden-layer network where the hidden layer acts as a fixed nonlinear feature map, and in which only the output weights are optimized. QELMs extend this concept to quantum systems, leveraging parametrized families of quantum states as feature maps. Notably, QELMs belong to the broader family of quantum feature extractors~\cite{bokov2026machine,zhang2026quantum}, for which formal quantum-classical separations are known~\cite{bokov2026machine}. 

For a given vector of classical inputs $\mathbf{u}$, a QELM prepares a state of the form $\rho(\mathbf{u}) = U_{\mathrm{RES}}U(\mathbf{u}) \rho_0 U^\dagger(\mathbf{u})U^\dagger_{\mathrm{RES}}$, where $\rho_0$ is an initial reference state, $U$ is the data encoding unitary and $U_{\mathrm{RES}}$ denotes a fixed, non-parametric reservoir evolution. More generally, one could also have 
\begin{equation}
    \rho(\mathbf{u}) = \left(\prod_{j=1}^\ell U_{\mathrm{RES}}U^{(j)}(\mathbf{u})\right) \rho_0 \left(\prod_{j=1}^\ell U_{\mathrm{RES}}{U^{(j)}}(\mathbf{u})\right)^\dagger \, ,
    \label{eq:layers_qelm}
\end{equation}
where $\{U^{(j)}\}$ are $\ell$ (potentially distinct) encoding layers. At each forward pass, outputs are collected from data-encoding states as expectation values over a fixed set of observables, namely $r_i(u) = \langle \hat{R}_i \rangle_{ \rho(u)} = \Tr[\hat{R}_i \rho(u)]$, where $\mathcal{R} = \{\hat{R}_i\}_{i=1}^{M}$ is a collection of Hermitian operators. In the following, we will use the vector of outputs $\mathbf{r}(u) \in \mathbb{R}^{M+1}$, where $r_0 = 1$ is used to incorporate a bias term. Once the outputs are collected, a classical linear function $f(u) = \mathbf{w} \cdot \mathbf{r}(u)$ is evaluated with respect to real weights $\mathbf{w} \in \mathbb{R}^{M+1}$. The weights of this linear layer are optimized offline after executing all circuits and collecting all expectation values from a given dataset. In contrast, the reservoir dynamical features remain unchanged throughout the whole model operation. These principles straightforwardly enable an effective use of cloud-based quantum resources, as -- contrary to common quantum neural networks based on parametrized quantum circuits -- the system does not require a constant and otherwise blocking exchange of information between the Quantum Processing Unit (QPU) and its classical counterparts. Furthermore, parametric circuit compilation can be leveraged to yield high sampling rates, particularly on superconducting processors~\cite{fischer2024dynamical,fischer2025large}. While similar properties are shared by the class of quantum kernel methods~\cite{havlivcek2019supervised,agliardi2025mitigating}, it is important to notice that only a linear number of forward evaluations (in terms of the dataset size) are required for QELMs, as opposed to the quadratic scaling of common fidelity-based quantum kernels. \\

\begin{figure}[t]
    \includegraphics[width=0.99\linewidth]{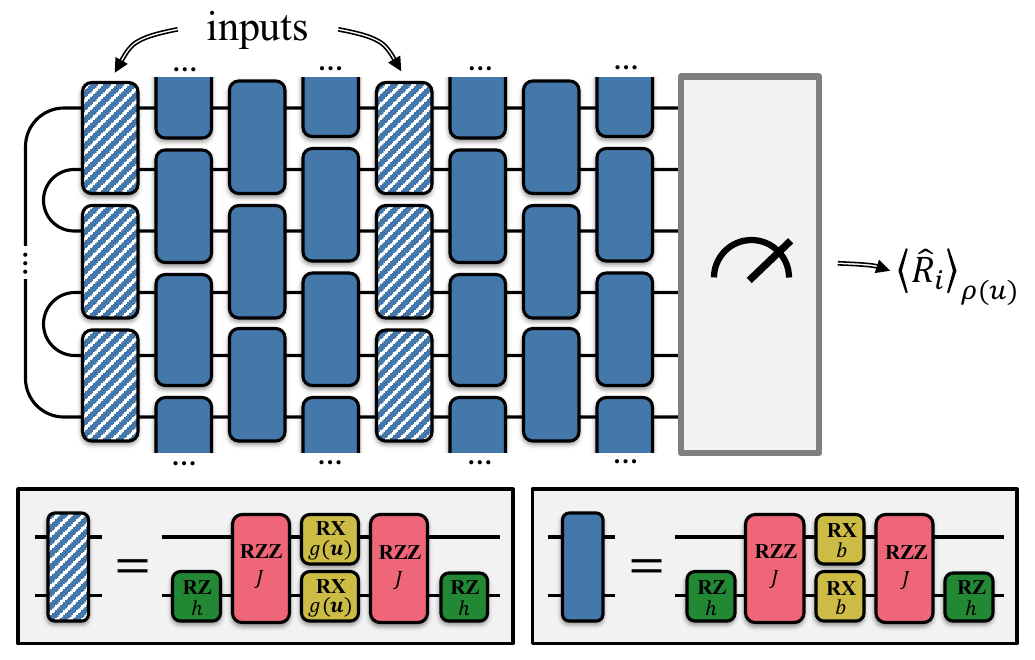}
    \caption{
        Schematic representation of the QELM architecture. The input vector $\mathbf{u}$ is injected through a sequence of encoding layers, where each two‑qubit block applies a data‑dependent transverse‑field kick with strength $b = g(u[i])$. These layers alternate with dynamics layers, in which the blocks implement fixed, randomly initialized kicked‑Ising evolutions governed by parameters $(J, h, b)$ that remain constant after initialization. Qubits are arranged in a ring topology and initialized in Bell pairs. After the full sequence of encoding and dynamical evolutions, expectation values of a set of observables $\mathcal{R}$ are measured from the final quantum state, forming the reservoir feature vector $\mathbf{r}(\mathbf{u})$. A classical linear readout layer combines these features to produce the final model output.} 
    \label{fig:QELM_schematic}
\end{figure}

The core design of our QELM is inspired by a class of circuits that simulate the time evolution of a specific kicked Ising model, which has recently been successfully implemented on superconducting quantum processors with more than 90 qubits~\cite{fischer2024dynamical}. This setup features a static Hamiltonian component, $\hat{H}_0$, and a periodically applied kick, $\hat{H}_K$, respectively given by
\begin{equation}
    \hat{H}_0 = J \sum_{i=0}^{N-2}\hat{Z}_{i}\hat{Z}_{i+1} + h \sum_{n=0}^{N-1}\hat{Z}_{i} \quad \text{and} \quad  \hat{H}_K = b \sum_{n=0}^{N-1}\hat{X}_{i} \,, 
\end{equation}
where $J$ denotes the spin-spin coupling, $h$ the longitudinal field, and $b$ the transverse kick strength. The resulting circuits adopt a brickwall layout of two-qubit blocks, each combining entangling gates and single-qubit rotations. Qubits are arranged in a periodic ring topology, initialized with adjacent pairs prepared in Bell states $(\ket{00}+\ket{11})/\sqrt{2}$.
A schematic representation of the QELM architecture is presented in Fig.~\ref{fig:QELM_schematic}.
Notably, this class of models exhibits a rich phenomenology as a function of the couplings and driving fields strength~\cite{bertini2025exactly}. As such, it represents an ideal, tunable test bed for our proposed benchmarking techniques. 

For an even $N$-qubit system, each layer consists of $N/2$ two-qubit blocks. Layers alternate between one \emph{encoding layer}, which inject input data leveraging the kick action, and three \emph{dynamics layers}. Notice that this construction is slightly more general (see Eq.~\eqref{eq:layers_qelm}) than the standard 1-layer QELM architecture, and akin to data reuploading schemes~\cite{perez2020data,schuld2021effect}. From a physically-inspired point of view, our model design may be interpreted as a repeated series of external input-dependent kicks interleaved within the system natural evolution. %
Future implementations of full QRC models could naturally leverage a similar structure by interleaving a variable stream of transverse-field inputs, measurements on visible nodes and reservoir evolution blocks. 

Each two-qubit block is parameterized by the three quantities $(J,h,b)$, where $J$ and $h$ determine the internal dynamics, while $b$ serves as the input encoding parameter for encoding blocks and acts as an additional dynamical parameter for dynamics blocks. At initialization, these parameters are independently drawn at random for each block from fixed normal distributions (e.g., $J \sim \mathcal{N}(J_0,\Delta J)$) and then kept constant. For encoding blocks, one input element is fed per block by setting $b=g(u^{(i)})$ where $u^{(i)}$ is the $i$-th element of the input vector $\mathbf{u}$ and $g(x)=a_{\text{in}}x$, where $a_{\text{in}}$ controls the input strength. The parameter distribution means and standard deviations, %
together with $a_{\text{in}}$, are treated as hyperparameters. These are not optimized during training but tuned upfront (see discussion below on optimal working points). Unless stated otherwise, we assume $N$ is even, so that $N/2$ is an integer, and we make use of $N/2$ layers to ensure a balance between input encoding and complex internal evolution. Each encoding layer can process up to $N/2$ distinct input elements simultaneously, resulting in a total potential input size of $\lceil N/8\rceil \times N/2 \in \mathcal{O}(N^2)$. The set $\mathcal{R}$ of measured observables includes weight-1 and weight-2 Pauli strings.

\section{Universal operating regime}
\label{sec:hyperparam_tuning}

\begin{figure*}[ht]
    \centering
    \includegraphics[width=\linewidth]{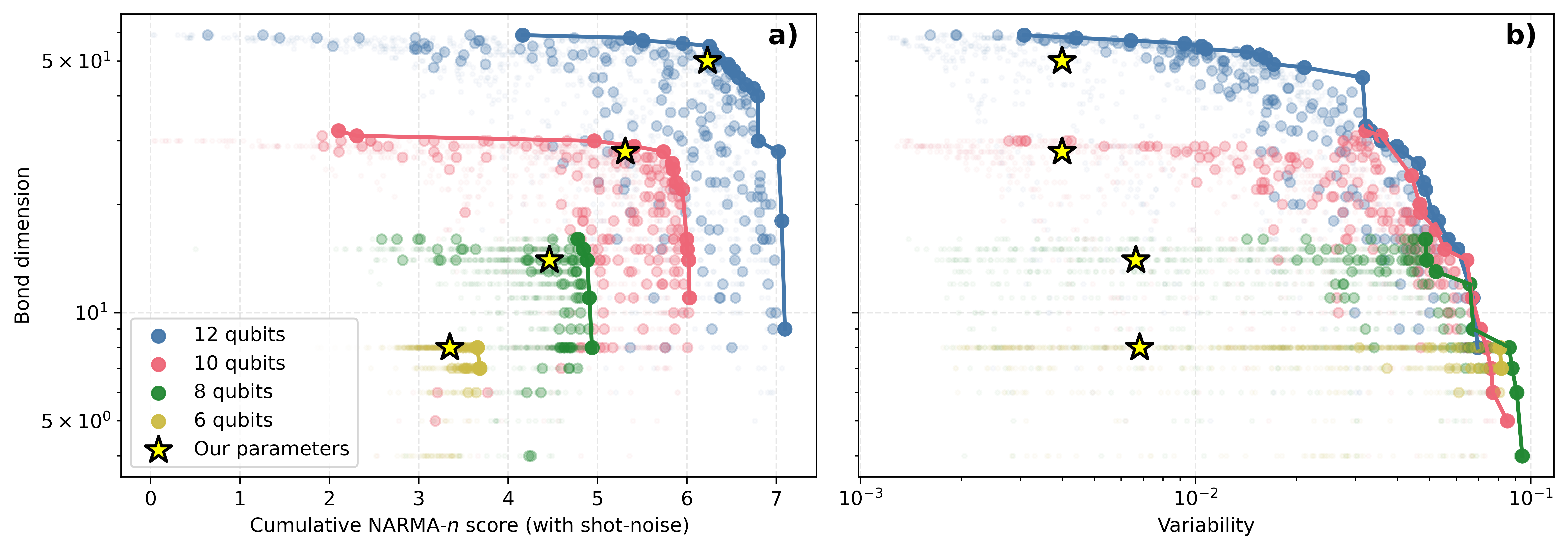}
    \caption{Pareto‑front projections obtained from multi‑objective hyperparameter optimization across QELM architectures of different sizes (6, 8, 10, and 12 qubits). Each dot represents a sampled hyperparameter configuration, colored by system size, and the solid lines indicate the corresponding Pareto fronts. Yellow stars mark the hyperparameters selected from the initial joint optimization (Table~\ref{tab:hyperparameters}). (a) Trade‑off between NARMA performance under realistic shot noise and bond dimension, illustrating how the chosen configuration consistently lies near the Pareto‑optimal region across model sizes. (b) Trade‑off between observable variability and bond dimension, showing that the selected operating point occupies a regime of high entanglement richness while preserving sufficient output variability to avoid over‑concentration. Together, these projections highlight the emergence of a universal operating regime that generalizes well across architectures.}
    \label{fig:qelm/pareto_projections}
\end{figure*}

Recent theoretical results formalized, through appropriate bounds~\cite{xiong2025fundamental}, a series of concentration behaviors which affect generic QELMs, conditioned on model properties such as Haar-expressivity, entanglement, global measurements, and noise. In practice, these effects appear as an exponential decay of the readout variances, evaluated over a given input distribution, as the size of the model grows. At the same time, QELM outputs can only be estimated in practice to finite precision via repeated quantum circuit sampling, typically with the well known inverse square root scaling of standard errors with the number of shots. As a result, when observables exhibit exponential concentration, distinguishing their true expectation values from a fixed mean requires exponentially many resources. Consequently, even though QELMs remain trainable offline via convex optimization, their generalization capabilities severely deteriorate as the measured features lose sensitivity to input data. At the same time, the absence of concentration effects in quantum models has also been consistently associated with an inherent classical simulabilty~\cite{cerezo2025does}, raising questions about the practical value of many of these solutions.

Initial works in the QRC literature have largely ignored the concentration issue, focusing instead on expressivity and complexity, and only recently this aspect has started to be thoroughly investigated from a mathematical point of view~\cite{hu_tackling_2023,xiong2025fundamental,xiong2025role}. However, ignoring these effects can lead to overly optimistic scalability claims. For instance, as highlighted by Xiong~et~al.~\cite{xiong2025fundamental}, the same mechanisms that induce non-trivial Fourier spectra in the model also exacerbate concentration. 

Hyperparameter choices strongly influence such trade-off between \emph{expressivity} (richness of the feature space), \emph{complexity} (as measured, e.g., by the amount of entanglement or non-stabilizerness exhibited by typical quantum states visited by the model) and \emph{resolvability} (accuracy under finite sampling). To this aim, we identify for our proposed QELM a suitable operating regime through a multi-objective tuning strategy informed by key performance metrics. These include: (i) the coefficient of determination ($R^2$), which measures the model’s explanatory power in time series prediction; (ii) the bond dimension of a Matrix Product State (MPS) representation of output states, used as an indicator of how efficiently such states can be simulated classically, and therefore useful for excluding parameter regimes that are unlikely to support any form of quantum advantage; and (iii) the output variance (hereafter referred to as \textit{variability}) over a uniform input distribution, which captures how strongly the observables respond to changes in the input. The latter is defined for each observable $\hat{R}_i \in \mathcal{R}$ as
\begin{equation}
\mathrm{Var}_{\mathbf{u}} \big[ r_i(\mathbf{u}) \big] = \mathrm{Var}_{\mathbf{u}\sim\mathrm{Unif(-1,1)}} \big[ \langle \hat{R}_i \rangle_{\rho(\mathbf{u})} \big].
\end{equation}
Further details on all metrics are reported in Appendix~\ref{app:methods}. We perform a multi-objective optimization over the hyperparameters $a_{\text{in}}$, $b_0$, $\Delta b$, $h_0$, $\Delta h$, $J_0$, and $\Delta J$ via \texttt{optuna}~\cite{akiba2019optunanextgenerationhyperparameteroptimization}, aiming at the optimal tradeoff between practicality and dynamical richness while avoiding regimes that are either trivial (e.g., weakly expressive) or overly chaotic (i.e., exhibiting poor processing capability under realistic shot noise). In all parameter searches we enforce $h>0.1$ to avoid the integrable limit: when the longitudinal field vanishes ($h=0$), the underlying Ising dynamics reduce to the transverse‑field Ising model, which is Jordan–Wigner solvable (i.e., effectively free‑fermionic)~\cite{suzuki2013quantum} and thus atypically simple compared to the generic regime we target. Furthermore, to prevent overfitting to a specific model size or depth, we perform the optimization using two distinct architectures, namely a 8-qubit model with 4 layers (one encoding and three dynamical ones), and a 10-qubit model with 8 layers, alternating one encoding and three dynamical steps. In the second model class, inputs are reuploaded twice.

After running the multiobjective optimization, we select one representative hyperparameter configuration lying near the Pareto front for the joint objectives (see Table~\ref{tab:hyperparameters}). The complete set of hyperparameter values selected is provided in Table~\ref{tab:hyperparameters}.
\begin{table}[ht]
\begin{tabular}{lrl}
 Hyperparameter & \multicolumn{2}{l}{Value} \\ \hline
 Input strength & $a_{\text{in}}$ &$= 0.2$ \\
 Transverse field & $b$ &$\sim \mathcal{N}(0.707, 0.031)$ \\
 Longitudinal field & $h$ &$\sim \mathcal{N}(0.683, 0.034)$ \\
 Coupling strength & $J$ &$\sim \mathcal{N}(0.237, 0.038)$ 
\end{tabular}
    \caption{Hyperparameter values used in all experiments. The transverse field $b$ is relevant only for dynamics blocks. The transverse field $b$, longitudinal field $h$ and coupling strength $J$ are sampled independently from normal distributions for each block (both dynamics and encoding blocks).}
    \label{tab:hyperparameters}
\end{table}
To validate that such choice generalizes well to other system sizes, we run additional, independent multi-objective optimizations for 6-, 8-, 10-, and 12-qubit models (with 4, 4, 8, and 8 layers respectively). For these, we target high bond dimension and output variability, as well as good NARMA performances both at infinite precision and under realistic statistical noise. While each of these optimization is performed separately and may overfit to the specific model size, our goal is to compare the location of our previously chosen optimal operating regime on the resulting Pareto fronts.

Interestingly, the results reported in Figure~\ref{fig:qelm/pareto_projections} show that our selected hyperparameters consistently lie, across all system sizes,  close to the Pareto-optimal trade-offs. In particular, the balance between bond dimension and NARMA performance under statistical noise (Fig.~\ref{fig:qelm/pareto_projections}a) demonstrates a good and transferable compromise between complexity, expressivity and robustness. Furthermore, Figure~\ref{fig:qelm/pareto_projections}b clearly shows how our choice of hyperparameters leans toward the richest available entanglement structure subject to a lower bound in observable variability. While, due to computational costs, we only focused on small‑to‑intermediate system sizes for all multi‑objective optimizations, our examination of how the Pareto‑optimal region transforms as we double the depth and increase qubit number already offers useful insights. Indeed, observing stable behavior (especially in terms of variability) over this range provides a meaningful proxy for much larger reservoirs, also in view of the local nature of the underlying spin‑model interactions. This intuition is further confirmed experimentally by the 24‑qubit NARMA study presented in the next Section, which provides a hardware‑level bridge towards the largest model instances.

Overall, our analysis provides evidence of a regime where hyperparameters identified based on paradigmatic benchmarks at small scales generalize well across tasks and reservoir sizes. Similar behavior has been reported in other architectures~\cite{kornjavca2024large,kobayashi2024feedback,solanki2025harnessing}, and is of crucial importance for subsequent realizations of QELMs beyond the regime of exact classical simulability.

\section{Applications and results}
\label{sec:results}

\begin{figure*}[ht]
    \centering
    \includegraphics[width=\linewidth]{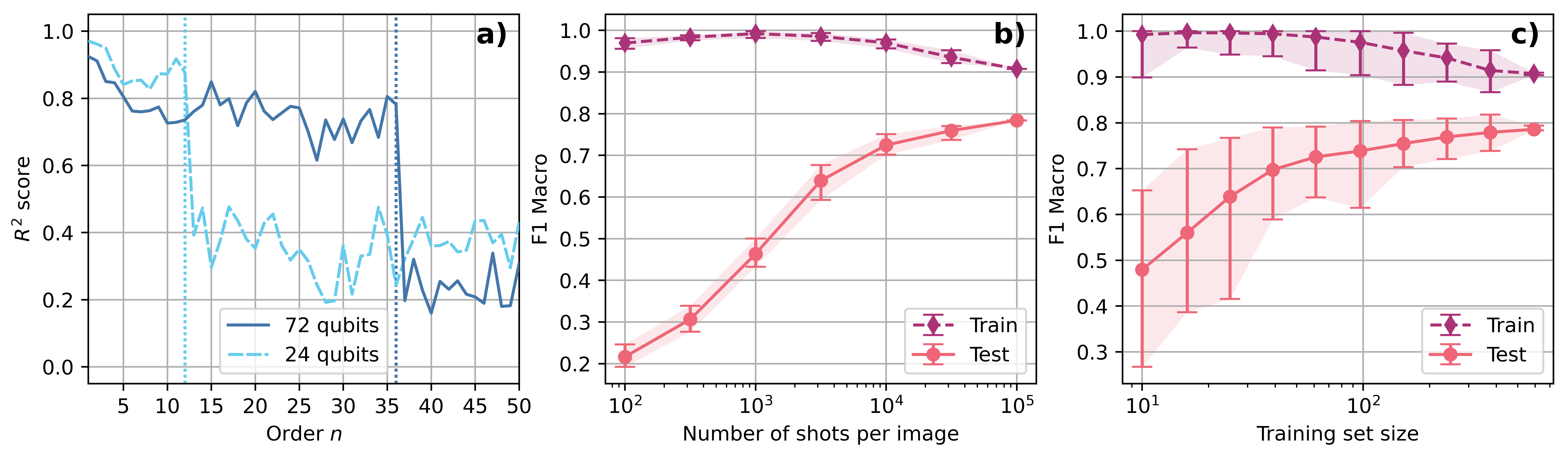}
    \caption{
        Hardware validation of large-scale QELM architectures on the \texttt{ibm\_quebec} processor.
        (a)~Performance on the NARMA‑$n$ benchmark for 24‑ and 72‑qubit QELMs executed on hardware. Here $n$ is the task order (target depends on the last $n$ outputs and specific input delays). $R^2$ is reported versus $n$; vertical dotted lines indicate the rolling‑window size $n=L=N/2$. The 72‑qubit model maintains accuracy to larger $n$, consistent with its greater effective memory. 
        (b)~Classification performance on the Landsat dataset for the 124-qubit QELM, shown as a function of the number of measurement shots used per input sample. Training and test F1 macro scores are displayed together with $95\%$ bootstrap confidence intervals, demonstrating systematic improvement in generalization as shot precision increases. 
        (c)~Learning curves for the same 124-qubit model, reporting the F1 macro score versus the size of the training set. Test performance improves monotonically with additional training data, while the training score remains consistently high, indicating that the linear readout operates far from saturation. Shaded regions and error bars again denote $95\%$ bootstrap confidence intervals.
    }
    \label{fig:qelm/hardware_experiments_results}
\end{figure*}

Having configured the QELM architecture and identified an optimal operating regime through numerical simulations, we now turn to its large-scale deployment on utility-scale superconducting quantum processors. Our goal in this section is to validate, on hardware, the effective short-term memory and predictive capability of large-scale QELMs, first on controlled nonlinear time-series benchmarks (NARMA-$n$) and then on a real-world classification task (Landsat). We refer the reader to Appendix~\ref{app:experiment_details} for a detailed description of the experimental setup.  \\

We evaluate NARMA-$n$ benchmarks on \texttt{ibm\_quebec} using 24- and 72-qubit QELM models. Here, the \emph{task order} $n$ denotes the memory horizon of the target recurrence: in NARMA-$n$, the reference output at time $t{+}1$ depends on the last $n$ outputs and on a multiplicative term involving inputs at delays $0$ and $n\!-\!1$ (see Appendix~\ref{app:methods}). In our QELM setup, sequences are processed via a rolling input window of length $L=N/2$ (one scalar per two-qubit block per input layer), so $L$ is the maximum effective delay the model can represent within a single forward pass. 

For each configuration, we use $T_{\mathrm{init}}{=}100$ warm-up steps (discarded), $T_{\mathrm{train}}{=}250$ training steps, and $T_{\mathrm{test}}{=}250$ test steps (consistent with Sec.~\ref{sec:hyperparam_tuning}). Performance is reported as the coefficient of determination $R^2$ versus the task order $n$. In this windowed (memoryless-per-window) regime, achieving high $R^2$ up to $n \approx L$ is a proxy for \emph{effective memory capacity}: the model retains and exploits information distributed across the $L$ most recent inputs.

The results, reported in Figure~\ref{fig:qelm/hardware_experiments_results}a, confirm that the model retains meaningful signal for task orders up to $n=N/2$, i.e., the size of the rolling input window.  This indicates that increasing the number of qubits improves the effective memory capacity of the system. %
Indeed, the 72-qubit system significantly outperforms the 24-qubit counterpart for $12 < n \leq 36$, consistently with its larger rolling window. For $n \leq 12$, we observe a slight drop in performance for the larger system, which we attribute to a more pronounced sensitivity to shot and hardware noise. In Appendix~\ref{sec:results/hardware_classification_task/QELM_multi_step}, we also describe a preliminary application of a 12-qubit QELM to the more challenging task of few-step-ahead prediction, using day-ahead hourly electricity price SPOT market data. \\ 

Besides time series, we also evaluate the model’s ability to perform supervised classification tasks based on the Statlog Landsat Satellite dataset from the UCI Machine Learning Repository~\cite{uci_statlog_landsat}, a prototypical ELM benchmark on real-world data. 

Experiments were conducted on the \texttt{ibm\_quebec} processor using a 124-qubit circuit architecture composed of 36 blocks, resulting in a total two-qubit gate depth of 91 and an overall count of 5084 two-qubit gates. As in Sec.~\ref{sec:hyperparam_tuning}, the circuit blocks follow a 1 ``input''-3 ``dynamics'' interleaved layout, with a total of 9 input layers and 27 dynamics layers. Each layer contains 62 two-qubit operations and, like previous experiments, all of these are used for data encoding in input layers. This configuration allows for the injection of up to 62 independent input features per layer. We consider both weight-1 and non-overlapping nearest neighbors weight-2 Pauli observables in arbitrary bases, which we retrieve via randomized Pauli measurements (i.e., local classical shadows)~\cite{huang2020predicting,dao2025dual} using the Qiskit POVM toolbox~\footnote{\url{https://qiskit-community.github.io/povm-toolbox/}}. 
Crucially, even at these scales our QELM models use the same set of hyperparameters identified in Section~\ref{sec:hyperparam_tuning}, without any additional task-specific tuning.

We create input vectors $\mathbf{u}$ of size 72 by stacking two copies of the 36 features of each Landsat dataset entry. We then feed entries $(u_l,u_{l+1},\ldots,u_{l+61})$ to the $l$-th encoding layer, shifting cyclically along $\mathbf{u}$ and the qubit ring. The dataset was sub-sampled to 860 images from the original 6435 available ones, and classification performance is then analyzed with respect the size of the training set and the number of measurement shots per image.

We initially benchmark the model performances making use only of single-qubit Pauli output data. In Fig.~\ref{fig:qelm/hardware_experiments_results}b--c we report \emph{F1 macro} scores, defined as the unweighted mean of the per-class F1 values, $\mathrm{F1} = 2(\mathrm{precision}\times\mathrm{recall})/(\mathrm{precision}+\mathrm{recall})$, a metric that is standard practice for multi-class, imbalanced datasets such as Landsat. Figure~\ref{fig:qelm/hardware_experiments_results}b shows that increasing the size of the training dataset improves generalization and class separation, reducing overfitting.
Additionally, Figure~\ref{fig:qelm/hardware_experiments_results}c shows that a larger number of shots further improves classification performance, indicating the model’s ability to learn robust decision boundaries.
In both cases, robust performances can be achieved under realistic resource constraints for large-scale QELM. Error bars are $95\%$ bootstrap confidence intervals, computed by resampling the training set in Fig.~\ref{fig:qelm/hardware_experiments_results}b, and the measurement shots in Fig.~\ref{fig:qelm/hardware_experiments_results}c, thus respectively capturing data‑ and sampling‑induced variability. Learning curves for additional metrics (accuracy, F1 weighted, and precision) are provided in Appendix~\ref{app:learning_curves}.

In the spirit of balancing measure concentration effects and achieving optimal information retrieval from large scale QELMs~\cite{martinez2023information,gross2026kernel}, we further extend our study to include weight-2 observables while leveraging \textit{eigentasks}, a framework originally introduced by Hu~et~al.~\cite{hu_tackling_2023}. Eigentasks (see also Appendix~\ref{app:methods}) are orthogonal model output functions that can be explicitly ranked by their Noise‑to‑Signal Ratio (NSR). They can be interpreted as identified directions in feature space that are maximally robust to shot noise. In the quantum setting, this corresponds to optimizing over linear combinations of measurement operators (POVM effects for our randomized Pauli measurements) or observables that maximize the so-called \emph{Resolvable Expressive Capacity} (REC).

\begin{figure}[t]
    \centering
    \includegraphics[width=0.99\linewidth]{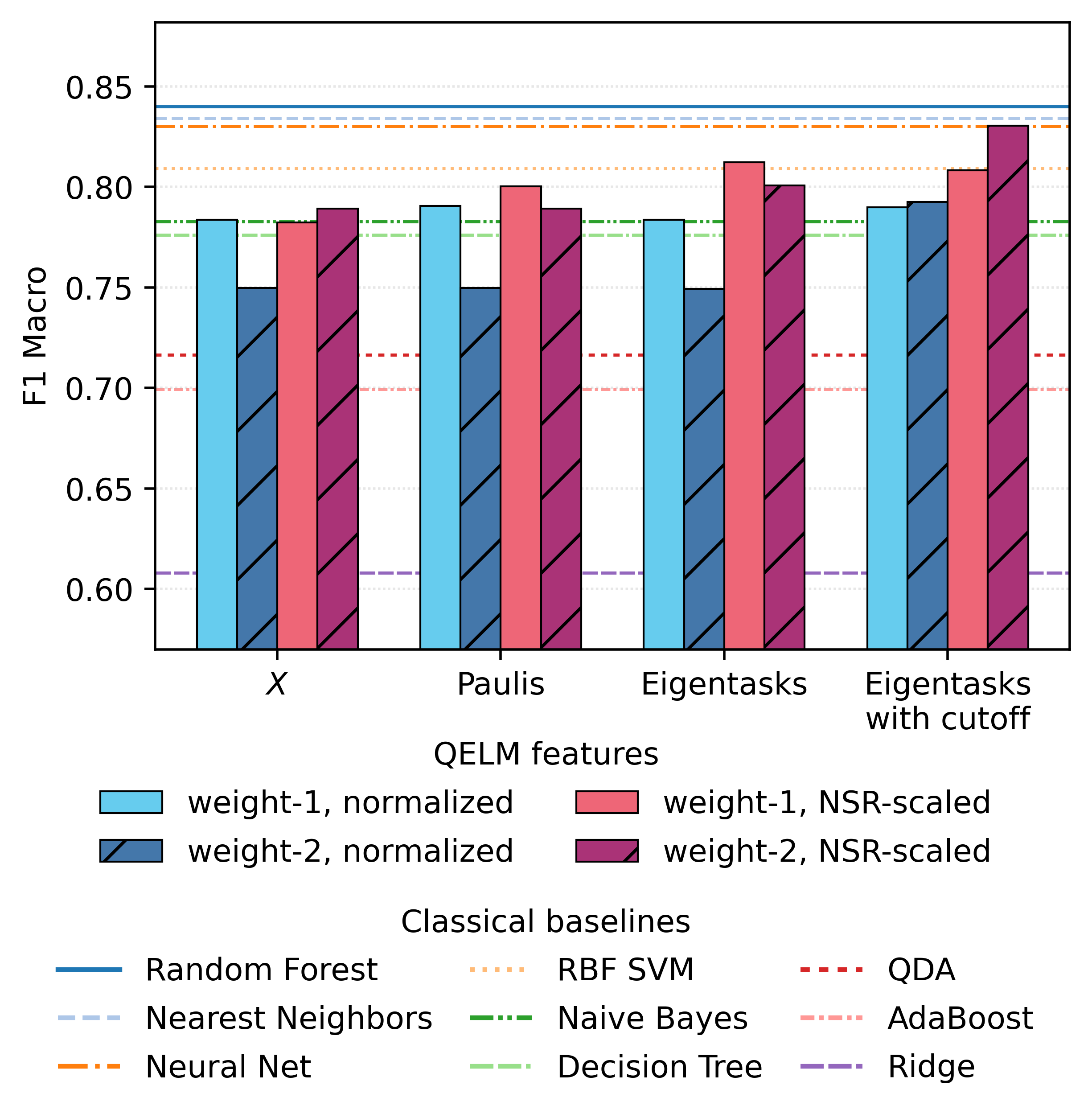}
    \caption{
        Classification performance (F1 macro) for the 124‑qubit Landsat experiment using different QELM readouts: single‑basis $X$, full Pauli, local eigentasks (ET), and ET with NSR‑based cutoff. Bars compare weight‑1 vs.\ weight‑2 features and unit-variance vs.\ signal‑aware scaling. Horizontal lines denote classical baselines trained directly on the raw features without tuning (see Appendix~\ref{app:results/classical_baselines} for configurations). Each baseline score is the average over 100 independent initializations and training runs.
    }
    \label{fig:pauli_vs_eigentask124}
\end{figure}

A central limitation of a global eigentask analysis is its scaling with system size $N$: the relevant operator space has dimension $4^N-1$, so both the covariance/Gram estimation and the ensuing generalized eigenproblem grow exponentially complex in $N$. This renders a full, global treatment unfeasible at large scales. For this reason, we introduce a novel \textit{local} approach (see Appendix~\ref{app:methods}) that operates on 1‑local and 2‑local subsets -- i.e., observables supported on one or two qubits -- whose computational cost scales only polynomially in $N$ and remains fully compatible with randomized Pauli measurements. We therefore obtain low‑weight, shot noise‑robust observables that we can compare to Pauli‑basis readouts using the data collected in our 124‑qubit experiments (see Figure~\ref{fig:pauli_vs_eigentask124}). 

Three main trends emerge for the Landsat classification task: first, scaling output features to reflect their variability -- via the NSR for eigentasks and the relative variability for $X$/Pauli features -- consistently outperforms unit‑variance normalization, as it prioritizes higher‑signal directions. Second, moving from weight‑1 to weight‑2 observables brings improvements for scaled eigentasks with the NSR cutoff applied, achieving the highest QELM score overall. For all other feature types, weight‑2 leads to lower scores at the same shot budget, which is consistent with the reduced variability of higher‑weight observables and with signal dilution across a larger measurement space. This effect is more pronounced for normalized features. Third, replacing raw Pauli features with local eigentasks consistently improves performance. For weight‑1 scaled readouts, our F1-macro increases from $0.782$ ($X$‑only) to $0.800$ (Pauli) and to $0.812$ (eigentasks). For weight‑2 scaled readouts, F1 improves from $0.789$ ($X$‑only$/ $Pauli) to $0.801$ (eigentasks), reaching $0.830$ with an NSR‑based cutoff that discards unresolved eigentasks. This confirms that ranking measurement directions by robustness to sampling noise -- and pruning those below resolution -- provides measurable benefits at utility scale. More details on the NSR cutoff sensitivity can be found in Appendix~\ref{sec:results/landsat_cutoff}, where we show that there exist a broad regime where performance is stable around the theoretically motivated threshold~\cite{hu_tackling_2023}.

For context, we also report classical baselines trained directly on the raw Landsat features (horizontal lines in Fig.~\ref{fig:pauli_vs_eigentask124}). We use standard \texttt{scikit‑learn} configurations without tuning, except for $k$‑nearest neighbours where we set $k$ to the number of classes ($k=7$), and for the MLP where we increased \texttt{max\_iter} to ensure convergence (Appendix~\ref{app:results/classical_baselines}). Interestingly, we observe that the QELM+Ridge pipeline (i.e., Ridge trained on quantum features) substantially outperforms a Ridge classifier trained directly on raw features, indicating -- consistently with similar recent results~\cite{zhang2026quantum} -- that the reservoir induces a better (i.e., more linearly separable) feature geometry. Second, our QELM readouts are competitive with several other classical methods (e.g., decision tree, RBF SVM, shallow neural network). For the Landsat task, $k$NN and Random Forest achieve slightly higher F1 macro; nonetheless, eigentask‑based readouts substantially narrow the gap and offer a principled path to improved noise robustness at scale. It is also worth noting that state-of-the-art classical models, such as Random Forest classifiers, can reach accuracies up to 0.91 on the full dataset~\cite{uci_statlog_landsat}, whereas our QELM reaches 0.87 accuracy on the reduced dataset considered here. For all classical baselines, each reported score is averaged over 100 independent initializations and training processes. Results for additional metrics (accuracy, F1 weighted, and precision) and their comparison to the same classical baselines are provided in Appendix~\ref{app:landsat_all_metrics}.

\section{Conclusions}
In this work, we presented a large‑scale implementation of QELM models on superconducting quantum processors, leveraging state‑of‑the‑art quantum simulation primitives and a versatile set of operational strategies designed to balance performance, expressivity, and robustness. We provided evidence for a universal operating regime in the model hyperparameters, which can be effectively characterized at relatively small system sizes and subsequently exploited for substantially more demanding experimental settings. %
We further introduced a novel local formulation of eigentasks analysis, leveraging fast randomized measurements, which remains computationally tractable at quantum‑utility scales while yielding measurable improvements over standard output feature selection methods. 

Using this framework, we demonstrated applications to representative regression and classification machine learning tasks where we achieved performance competitive with classical learning techniques. While our framework does not yet exhibit genuine memory capacity, all of the model set-up techniques and design concepts developed here extend, in principle, to a larger class of reservoir-based methods (e.g., fully dynamical QRC). We expect that a successful implementation of the latter at scale could substantially narrow, and eventually reverse, the gap with respect to leading classical techniques, and represent therefore a natural next target for future research. %

Overall, our results add physical computing methods to the growing body of theory-informed empirical work which showcases a viable, pre‑fault‑tolerant quantum machine learning approach that is able to operate effectively at the one hundred qubit scale. Crucially, the availability of such models, combined with continuing improvements in reliability and speed of quantum processors, opens the possibility of concrete tests on real-world data at meaningful sizes. While the quest for provable quantum advantage stands, and might remain open beyond the near term, these demonstrations represent necessary %
steps towards such a long sought after goal.

\section*{Acknowledgments}
We thank the members of the Quantum Computing for Sustainability Working Group for stimulating discussions and useful insights. We acknowledge use of access provided by Plateforme d' Innovation Numérique et Quantique ($\mathrm{PINQ}^2$) to the \texttt{ibm\_quebec} device, which was used to enable algorithm development and hardware executions in this project. Timoth\'{e}e Dao and Francesco Tacchino acknowledge support from the Swiss National Science Foundation through grant 225229 (RESQUE). Ege Yilmaz and Stefan Woerner acknowledge support from the Swiss National Science Foundation through grant 214919.

\bibliography{references}

\clearpage
\onecolumngrid
\appendix

\section{Methods}
\label{app:methods}

\textbf{Rolling window inputs.} To process time series data through our (strictly speaking memoryless) QELM model, we adopt a time windowing approach. Specifically, for any input sequence $\{u_t\}$ of length $T$ and a $N$-qubit reservoir (assuming $N$ is even) we set a window size $L = N/2$ and we construct vectors of the form 
\begin{equation}
    \mathbf{u}_t = (u_t, u_{t-1}, \dots, u_{t-(L-1)})^\top \in \mathbb{R}^L.
    \label{eq:window}
\end{equation}
Each window $\mathbf{u}_t$ is then encoded and evolved by the QELM by feeding it entirely to each input layer (one entry per pair of qubits). At each layer, entries of $\mathbf{u}_t$ are shifted cyclically by four positions along the qubit ring. Crucially, since no reservoir state is carried across subsequent windows beyond $\mathbf{u}_t$ itself, the memory capacity of the model is upper-bounded by $L = N/2$. \\

\textbf{NARMA time series.} The Nonlinear Auto-Regressive Moving Average (NARMA) task of order $n$~\cite{kubota2023temporal,martinez2021dynamical,murauer2025feedback} is defined by the reference output sequence
\begin{equation}
    y_{t+1} = 0.2 y_t + 0.04 \left( \sum_{i=0}^{n-1} y_{t-i} \right) + 1.5 u_{t-(n-1)}u_{t} + 0.001,
\end{equation}
where the input sequence $u_t$ is sampled from $\mathrm{Unif}(0,0.5)$. In this recurrent relation, each reference output depends on all previous outputs (and thus past inputs), although the influence of earlier outputs and inputs diminishes for delays greater than $n$.

In all our benchmarks and experiments, we feed transformed inputs $\tilde{u}_t = 2u_t - 1 \sim\mathrm{Unif}(-1,1)$ to the QELM model using independently sampled length-$T_\mathrm{tot}$ sequences. More specifically, we sample $T_{\mathrm{tot}} = T_{\mathrm{init}} + T_{\mathrm{train}} + T_{\mathrm{test}} + (L-1)$ inputs. From the last $T_{\mathrm{train}} + T_{\mathrm{test}} + (L-1)$ values, we form a total of $T_{\mathrm{train}} + T_{\mathrm{test}}$ windows $\mathbf{u}_t$ according to Eq.~\eqref{eq:window} for $t \in \{1,\ldots,T_{\mathrm{train}} + T_{\mathrm{test}}\}$. The initialization segment of length $T_{\mathrm{init}}$ is not fed to the model, as no reservoir state is carried over. \\

\textbf{Model training.} For the hyperparameter tuning study described in Sec.~\ref{sec:hyperparam_tuning}, we collect the average values of single-qubit Pauli operators ($X$, $Y$, $Z$) and nearest-neighbor two-qubit correlators ($XX$, $YY$, $ZZ$) across the whole system. In larger experiments, see Sec.~\ref{sec:results}, we measure instead the complete set of nearest-neighbor two-qubit Pauli operators. In all cases, the set of output expectations forms a feature vector $\mathbf{r}_t \in \mathbb{R}^{M}$, where $M=|\mathcal{R}|$. By stacking over the time axis, we form the matrices $\mathbf{R}_{\mathrm{train}} \in \mathbb{R}^{T_{\mathrm{train}}\times M}$ and $\mathbf{R}_{\mathrm{test}} \in \mathbb{R}^{T_{\mathrm{test}}\times M}$. Finally, we train a linear readout layer $\hat{y}_t = \mathbf{w} \cdot \mathbf{r}_t + w_0$ by least squares on the training set, minimizing the mean squared error $\mathrm{MSE} \propto  T^{-1} \sum_t \norm{{y}_t- \mathbf{w} \cdot \mathbf{r}_t}_2^2$, with some regularization term $\lambda \norm{\mathbf{w}}_2^2$. For each task, the model is then evaluated on test data by the coefficient of determination, defined for scalar outputs as $R^2 = 1-\text{MSE}/\text{Var}_t[y_t]$. This quantifies the proportion of variance in the dependent variable that is accounted for by the model, thus serving as a measure of the model’s explanatory power. \\

\textbf{Hyperparameter search.} As mentioned in Sec.~\ref{sec:hyperparam_tuning}, for each candidate hyperparameter configuration we evaluate the NARMA score and typical bond dimension for 8- and 10-qubit models.
These objectives are optimized jointly so that the selected parameters generalize across different system sizes rather than overfitting to a specific architecture. For an $n$-qubit reservoir (processing an input of length $n/2$), the two metrics are defined as follows: (i) the \textit{NARMA score} corresponds to the cumulative $R^2$ over NARMA tasks of orders $1$ through $\lfloor n/2 \rfloor$. Expectation values are obtained from noiseless statevector simulation, after which Gaussian noise $\mathcal{N}(0,10^{-3/2})$ can be added to emulate shot noise. We use 100 warm-up steps, 250 training steps, and 250 test steps. This metric captures the effective processing capacity of the reservoir under realistic sampling noise, and penalizes regimes suffering from concentration effects. (ii) To estimate the typical model \emph{bond dimension}, we simulate the exact output state produced by a NARMA input sequence (scaled by $a_{\text{in}}$) and determine the smallest MPS bond dimension required to approximate this state with high accuracy. Specifically, we choose the minimal bond dimension for which the MPS reconstruction achieves fidelity $F \ge 1 - 5\times 10^{-4}$. This metric captures the entanglement generated by the reservoir dynamics and excludes parameter regimes that yield states amenable to efficient classical simulation.

For every hyperparameter setting we average all metrics over 10 random realizations, where each realization resamples the block parameters $(b,h,J)$ and draws a new NARMA input sequence. \\

\textbf{Classification benchmark.} The Landsat dataset consists of multispectral satellite images collected by the Landsat program. Each input sample corresponds to a $3 \times 3$ pixel neighborhood, where each pixel is described by four spectral bands, yielding 36 features per input sample. The classification task involves assigning each neighborhood to one of six land-cover categories.

Within our QELM framework, each input vector is encoded and processed similarly to the NARMA case. Outputs from the quantum model are then processed by a linear readout layer trained with an $\ell_2$-regularized least-squares objective, i.e.\ ridge regression on one-hot class labels. This readout produces class scores as linear functions of the reservoir features, with the final prediction obtained via an $\arg\max$ over these scores. By not adding further classifier assumptions, this setup makes performance depend directly on the feature structure and separability induced by the quantum reservoir and allows us to assess it expressive capacity~\cite{hu_tackling_2023}. \\

\textbf{Local eigentasks analysis.} When physical readouts are stochastic under finite sampling, the set of functions that can be robustly approximated is, in general, limited. The \emph{resolvable expressive capacity} (REC) formalizes this concept by identifying \emph{eigentasks} -- orthogonal functions ranked by their noise‑to‑signal ratio (NSR) -- as the directions that saturate REC~\cite{hu_tackling_2023}. If $\{\tilde{p}_k(u)\}_{k=1}^K$ are stochastic features obtained for input $u$ with $S$ measurement shots and $\mathbf{p}(u)=\EE[\tilde{\mathbf{p}}(u)]$ denotes the corresponding vector of expectation values, eigentasks are defined by the solution of the generalized eigenproblem
\begin{equation*}
    \mathrm{V}\,\mathbf{h}^{(l)} = \beta_l^{2}\,\mathrm{G}\,\mathbf{h}^{(l)},
\end{equation*}
as $y^{(l)}(u)=\mathbf{h}^{(l)}\!\cdot\!\mathbf{p}(u)$ with NSR $\beta_l^{2}$. Here $\mathrm{V}$ and $\mathrm{G}$ are the covariance and signal Gram matrices
\begin{equation*}
    \mathrm{V} = \EE_u\!\left[ \mathrm{Cov}\big(\tilde{\mathbf{p}}(u)-\mathbf{p}(u)\big) \right], 
    \quad 
    \mathrm{G} = \EE_u\!\left[ \mathbf{p}(u)\mathbf{p}(u)^{\!\top} \right].
\end{equation*}
In the context of our QELM realization, $\{p_k(u)\}_k$ are in general Born probabilities of a POVM $\{\hat{M}_k\}_k$ on state $\rho(u)$; the corresponding eigentask can then be written as
\begin{equation*}
\begin{split}
    y^{(l)}(u) &= \sum_k h^{(l)}_k\, \Tr[\hat{M}_k \rho(u)]
    = \Tr\!\left[\hat{R}_l \rho(u)\right], \\
    \hat{R}_l &= \sum_k h^{(l)}_k \hat{M}_k.
\end{split}
\end{equation*}
From  this, we see that eigentasks can be interpreted as defining adaptive measurement bases, ranked by robustness to shot noise. Following~\cite{hu_tackling_2023}, a standard resolvability criterion retains only eigentasks with $\beta_l^2 < S$.

Since a global eigentasks computation is exponentially complex in system size, we introduce a \emph{local} variant that computes eigentasks on fixed‑size subsets of qubits.  
We consider two collections of disjoint subsets,
\[
    \mathcal{S}_1 = \{ \{0\},\dots,\{N-1\} \},
    \quad
    \mathcal{S}_2 = \{ \{0,1\},\{2,3\},\dots \},
\]
corresponding to weight‑1 and weight‑2 local neighborhoods respectively. For each subset $S\in\mathcal{S}$ we solve the local generalized eigenproblem using randomized single-qubit Pauli measurement data restricted to $S$, yielding operators $\{\hat{R}^{(S)}_l\}_l$.  
Collecting across all subsets produces the candidate set
\begin{equation}
    \mathcal{R} = \bigcup_{S\in\mathcal{S}} \{\hat{R}^{(S)}_l\}_l ,
\end{equation}
where all operators are linearly independent.
This approach retains scalability while optimizing over all low‑weight observables supported on each subset. The approach offers two advantages: (i) low‑weight observables mitigate concentration effects~\cite{xiong2025fundamental}; (ii) randomized Pauli measurements broaden the accessible operator span beyond a single basis, improving resolvability at fixed shot budgets. The resulting local eigentask features are then fed to linear regressors/classifiers. 

Following~\cite{hu_tackling_2023}, an eigentask is considered \emph{resolvable} if its NSR satisfies $\beta_l^2 \le S$, where $S$ is the shot budget. We therefore define a cutoff‑filtered set
\begin{equation}
    \mathcal{R}_{\mathrm{NSR\text{-}cutoff}} = \{\, \hat{R}_l \in \mathcal{R} : \beta_l^2 \le S \,\}.
\end{equation}
Eigentasks with $\beta_l^2 > S$ are treated as unresolved, since their noise dominates their signal. Sensitivity to this cutoff is examined in Appendix~\ref{sec:results/landsat_cutoff} (Fig.~\ref{fig:eigentask124_vs_cutoff}).

To study how feature scaling interacts with eigentask structure, we compare several strategies. Unit‑variance normalization is defined by
\begin{equation}
    \hat{r}_{i,\mathrm{unit}}(u) = \frac{\hat{r}_i(u)-\mu_i}{\sigma_i},
\end{equation}
with $\mu_i=\EE_u[\hat{r}_i(u)]$ and $\sigma_i^2 = \mathrm{Var}_u[\hat{r}_i(u)]$. For eigentasks, we introduce an \emph{NSR‑aware scaling} that amplifies highly resolvable directions,
\begin{equation}
    \hat{y}^{(l)}_{\mathrm{NSR}}(u) = \frac{\beta_l^{-1}}{\max_{l'\in\mathcal{L}}\beta_{l'}^{-1}} \,\hat{y}^{(l)}_{\mathrm{unit}}(u),
\end{equation}
where $\mathcal{L}$ is the set of retained local eigentasks. For arbitrary observables $\hat{R}_i\in\mathcal{R}$ we also consider signal‑based scaling,
\begin{equation}
    \hat{r}_{i,\mathrm{signal}}(u) = \frac{\sigma_i}{\max_{i'\in\mathcal{R}}\sigma_{i'}} \,\hat{r}_{i,\mathrm{unit}}(u).
\end{equation}
Combining (i) the presence/absence of the NSR cutoff with (ii) either unit‑variance or NSR‑aware scaling, we evaluate all four regimes for local eigentasks. This allows a systematic assessment of how resolvability filtering and scaling jointly influence downstream model performance.

\section{Device and execution details}
\label{app:experiment_details}

\subsection{Quantum processor}

All hardware experiments in this work were carried out on an IBM Quantum Heron processor, specifically the \texttt{ibm\_quebec} device, accessed via the IBM Quantum Platform. 

\begin{figure}[ht]
    \centering
    \includegraphics[width=0.70\linewidth]{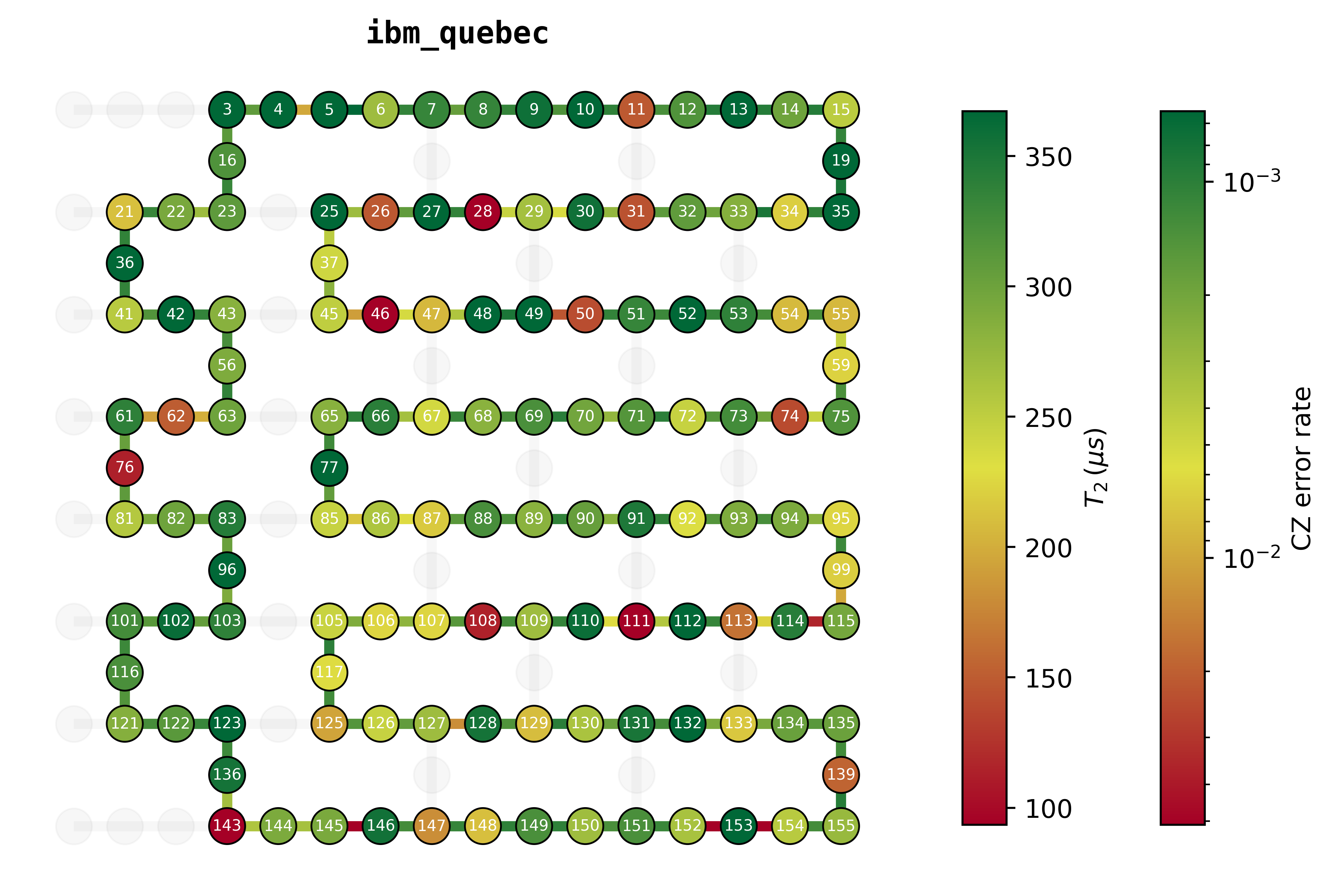}
    \caption{Qubit layout of the 124‑qubit ring used on \texttt{ibm\_quebec}.
    The heavy‑hexagonal lattice is shown in grey; the selected 124‑qubit cycle is highlighted. Node color encodes $T_2$ times, and edge color shows the calibrated $CZ$ error rates of the device.}
    \label{fig:ibm_quebec_124}
\end{figure}

The \texttt{ibm\_quebec} processor comprises 156 fixed-frequency transmon qubits arranged in a heavy-hexagonal lattice (see Fig.~\ref{fig:ibm_quebec_124} for the layout). The device exhibits a median relaxation time of $T_1 = 248.6 \, \unit{\micro\second}$ and a median dephasing time of $T_2 = 307.5 \, \unit{\micro\second}$. Quantum circuits are compiled into alternating layers of parallel single-qubit gates and two-qubit entangling controlled-$Z$ ($CZ$) gates. Single-qubit operations use $\sqrt{X}$-pulses ($SX$) and virtual $RZ$ gates~\cite{PhysRevA.96.022330}. The median infidelities are \num{1.63E-04} for $SX$, \num{1.53E-03} for $CZ$ and \num{4.52E-03} for the readout.

Circuits were transpiled to the device’s ISA using Qiskit (optimization level~3), with dynamical‑decoupling sequences (\texttt{XpXm}) applied throughout. Randomized, single‑qubit measurement bases were used for all readout experiments, allowing reconstruction of both single‑qubit and nearest‑neighbour two‑qubit Pauli operators.

For each experiment, we selected a connected ring subgraph of the heavy‑hex topology (sizes 24, 72, and 124). The 124‑qubit configuration used for the Landsat classification task is shown in Fig.~\ref{fig:ibm_quebec_124}.

\subsection{Execution summary}

A consolidated description of all hardware runs is provided in Table~\ref{tab:experiment_details}. Shot budgets refer to the number of measurement shots per input sample, and ``2q‑depth’’ denotes the depth of native $CZ$ layers after transpilation. All experiments used the Qiskit IBM Runtime \texttt{Sampler} primitive.

\begin{table}[h]
\centering
{\setlength{\tabcolsep}{7pt}
\begin{tabular}{llllllll}
\hline
 Experiment & Type & Device & Date & $N$ qubits & 2q‑gates & 2q‑depth & Shots/sample \\
\hline
 NARMA‑24 & Regression & \verb|ibm_quebec| & 2026‑01‑27 & 36  & 372  & 31 & \num{99999}  \\
 NARMA‑72 & Regression & \verb|ibm_quebec| & 2026‑02‑05 & 72  & 2196 & 61 & \num{540000} \\
 LANDSAT‑124 & Classification & \verb|ibm_quebec| & 2026‑01‑15 & 124 & 5084 & 91 & \num{99999}  \\
\hline
\end{tabular}}
\caption{Hardware resources for the experiments.
All circuits were compiled to ISA-level operations with optimization level~3 and executed with randomized local measurements. Reconstructed observables include $P_i$ and $P_i P'_{i+1}$ for $P\in\{X,Y,Z\}$.}
\label{tab:experiment_details}
\end{table}

\subsection{Measurement model and linear readout as POVM post‑processing}

It is convenient to formalize the readout as a post‑processing over a fixed Positive Operator‑Valued Measure (POVM) $\{\hat{M}_k\}_k$ implemented by randomized local measurements. For an input $\mathbf{u}$ that prepares $\rho(\mathbf{u})$, the observed probabilities are $p_k(\mathbf{u})=\mathrm{Tr}[\hat{M}_k \rho(\mathbf{u})]$. A linear readout with weights $\{w_k\}$ then computes
\begin{equation}
    f(\mathbf{u})=\sum_k w_k\, p_k(\mathbf{u})=\mathrm{Tr}\!\left[\hat{O}_w\, \rho(\mathbf{u})\right],
    \qquad
    \hat{O}_w \coloneqq \sum_k w_k\, \hat{M}_k,
\end{equation}
i.e., optimizing a linear readout is equivalent to selecting an effective observable $\hat{O}_w$ in the span of the POVM effects. This point of view matches our experimental workflow: (i) fix measurement settings; (ii) collect samples; (iii) train the readout offline.

\section{QELM extension for Multi-Step Energy Price Forecasting}
\label{sec:results/hardware_classification_task/QELM_multi_step}

We demonstrate a preliminary application of our QELM architecture to the challenging problem of 4-step-ahead SPOT market energy price prediction using day-ahead hourly electricity market data from January to July 2015. As in the main text, the quantum reservoir consists of a 12-qubit kicked Ising circuit initialized with Bell pairs, implementing a parametrized unitary evolution $U(\theta)$ where input signals modulate rotation angles via encoding blocks. The circuit employs a brickwork entanglement pattern and measures weight-1 and (nearest-neighbor) weight-2 Pauli observables, which serve as quantum features for downstream Ridge regression.

Three temporal features were engineered: normalized price, price differential (capturing velocity), and hour-of-day (encoding circadian patterns). Input preparation exploits a key architectural insight: while classical Echo State Networks (ESNs) accept full rolling windows, QELM circuits have fixed parameter slots. We extract the most recent $N/2$ timesteps from a larger historical window (12 hours), where $N$ is the qubit count, leveraging quantum entanglement for temporal memory rather than explicit history encoding. This design enables flexible window sizes without circuit modification. To accomodate three time series as input in our QELM model, we adopt a repeated pattern of 3-input and 3-dynamics layers (1 input layer per input time series for each repetition of the pattern).

\begin{figure}[ht]
    \centering
    \includegraphics[width=0.68\linewidth]{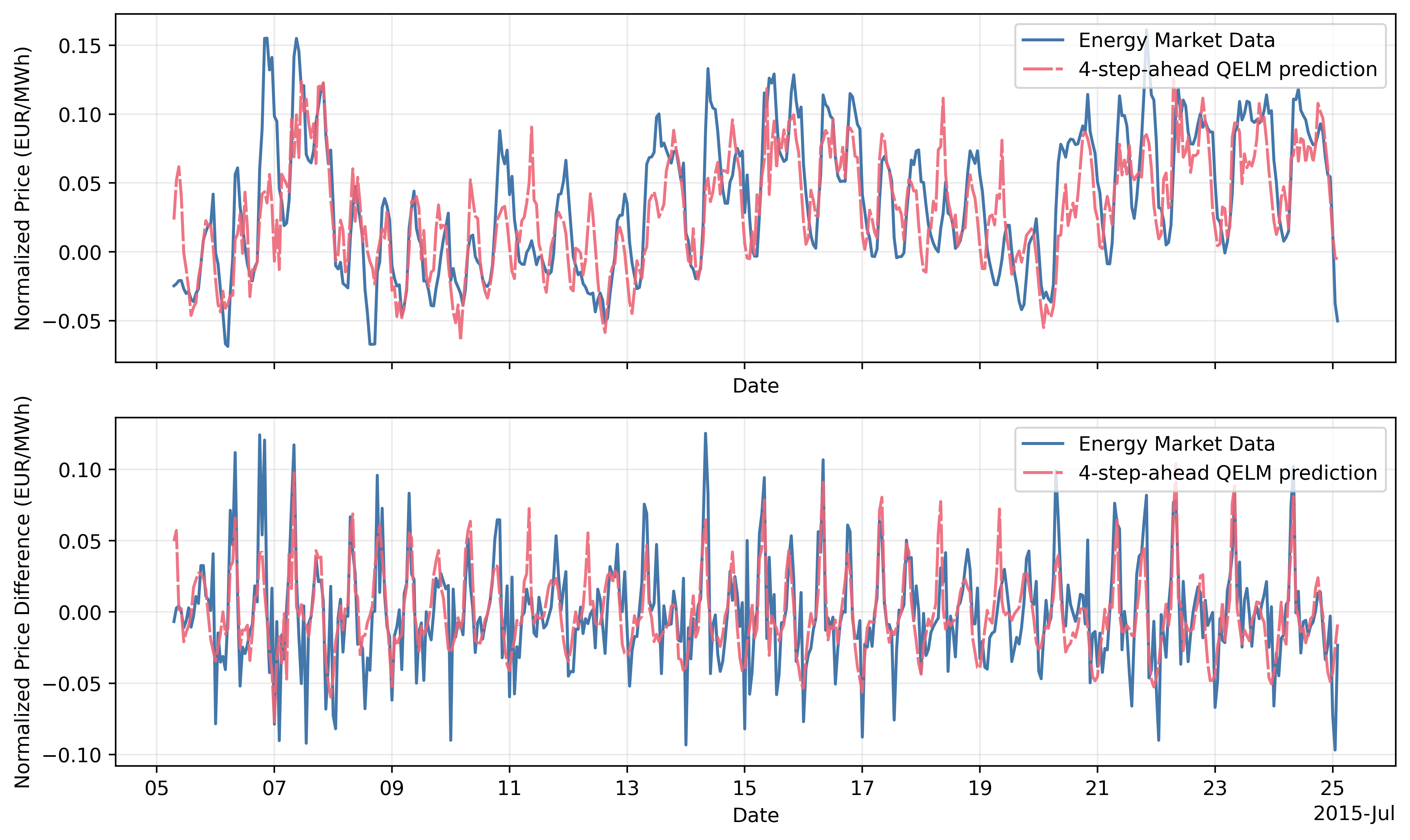}
    \caption{
        Four‑step‑ahead forecasting of energy‑market time series using a 12‑qubit QELM.
        Top: normalized electricity price and corresponding QELM prediction.
        Bottom: normalized price difference (temporal derivative) and QELM prediction. The model reproduces the dominant temporal structure of the market signal over several consecutive days.
    }
    \label{fig:energy_market_predictions}
\end{figure}

Using a Qiskit Aer simulator backend \cite{javadiabhari2024quantumcomputingqiskit} with \num{4785} temporal samples (90\% training, 10\% test), the QELM achieved competitive performance: MSE of \num{1.08e-3} for price prediction and \num{7.49e-4} for price differentials, matching a 400-neuron ESN baseline. Notably, QELM demonstrated 6\% superior performance on derivative prediction and achieved 76.3\% directional accuracy. Figure~\ref{fig:energy_market_predictions} displays the normalized prediction performance for both the 4‑step‑ahead price (top panel) and the corresponding price difference (bottom panel). The QELM predictions track the temporal structure of the target market signal over multi‑day horizons, capturing the dominant oscillatory components and short‑term intra‑day variability. %
Future work will explore larger circuits (utility scale) on quantum hardware, optimized entanglement structures, and extended prediction horizons. %

\section{Additional Landsat results}

\subsection{Learning Curves}
\label{app:learning_curves}

\begin{figure}[ht]
    \centering
    \includegraphics[width=0.8\linewidth]{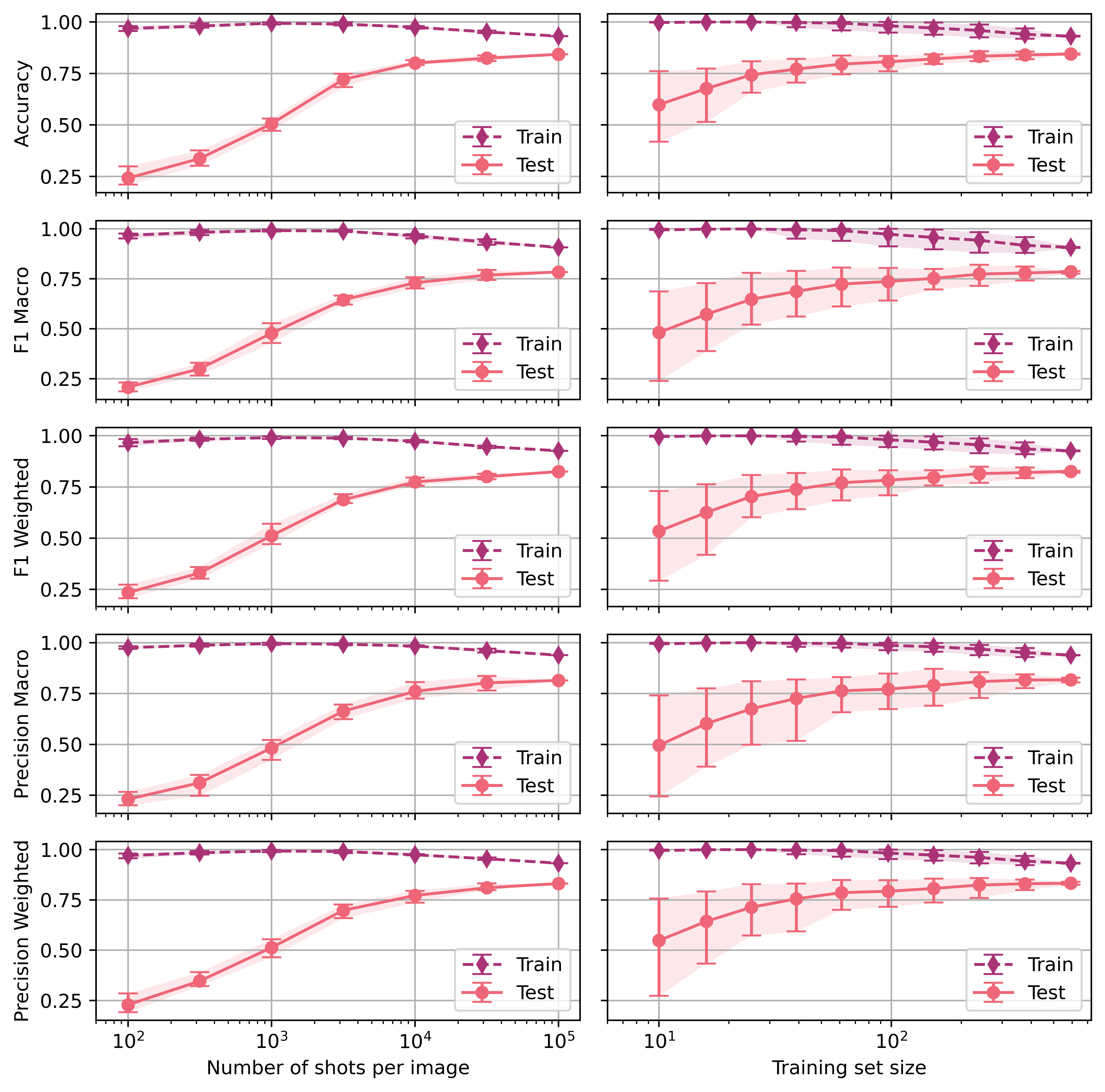}
    \caption{
        Learning curves for the 124‑qubit Landsat experiment.
        Left column: performance versus number of measurement shots per sample.
        Right column: performance versus size of the training set.
        Metrics reported include accuracy, F1 macro, F1 weighted, precision macro, and precision weighted. Solid lines denote mean values and shaded regions show 95\% bootstrap confidence intervals.}
    \label{fig:learning_curves}
\end{figure}

In this appendix we report the full set of learning curves associated with the 124‑qubit Landsat experiments discussed in Sec.~\ref{sec:results}. Figure~\ref{fig:learning_curves} shows the evolution of multiple performance metrics -- accuracy, F1 macro, F1 weighted, precision macro, and precision weighted -- as a function of (left column) the number of shots used per input sample and (right column) the number of training samples.

As expected for linear readouts trained on noisy quantum features, all metrics improve monotonically with the number of shots, reflecting a reduction in the effective estimator variance. The curves also display the characteristic generalization gap between train and test performance, which progressively closes with increased training‑set size. Importantly, all metrics exhibit smooth scaling behaviour consistent with the trends observed in the F1‑macro learning curve of Fig.~\ref{fig:qelm/hardware_experiments_results}(c), thus confirming that the qualitative conclusions drawn there extend across standard classification metrics.

\subsection{Pauli and eigentask readouts at 124 qubits}
\label{app:landsat_all_metrics}

\begin{figure}[ht]
    \centering
    \includegraphics[width=\linewidth]{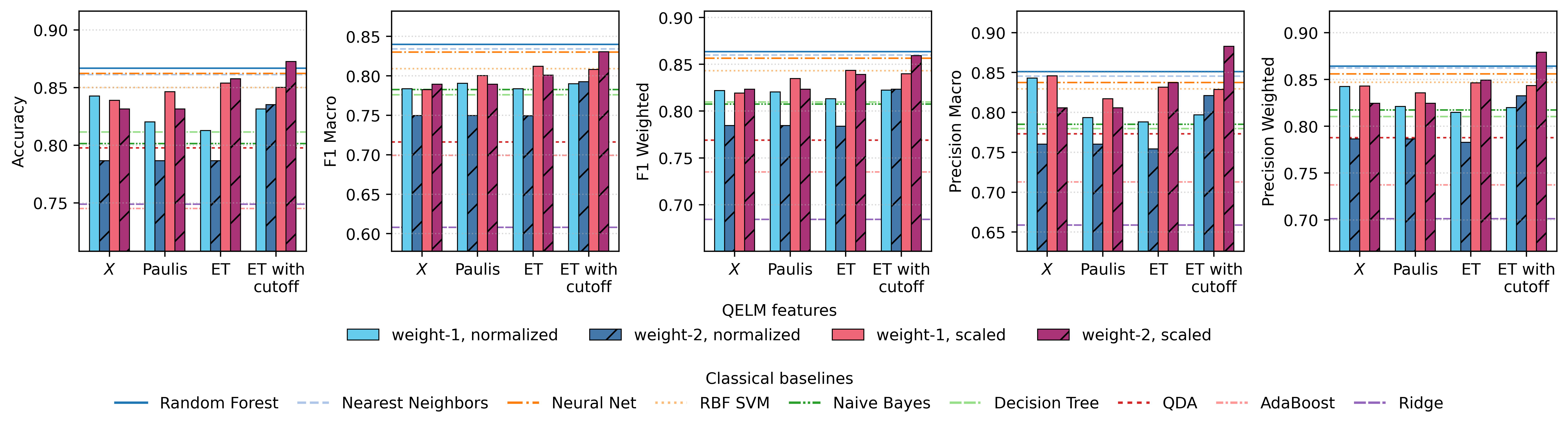}
    \caption{Comparison of Pauli and eigentask readouts (124‑qubit Landsat experiment).
    Accuracy, macro‑/weighted‑F1, and macro‑/weighted‑precision are reported for $X$‑only, full Pauli, local eigentasks (ET), and ET with an NSR‑based cutoff. Two feature scalings are shown: unit‑variance normalization (solid) and signal‑aware/NSR scaling (hatched). Horizontal lines denote classical baselines trained on raw features (see Appendix~\ref{app:results/classical_baselines}). Each classical baseline score is the mean over 100 independent initializations and training runs.
    }
    \label{fig:eigentask_vs_pauli_all_metrics}
\end{figure}

Figure~\ref{fig:eigentask_vs_pauli_all_metrics} compares several readout choices -- $X$‑basis, full Pauli, local eigentasks (ET), and eigentasks with an NSR cutoff -- across five standard metrics. Results are shown for both weight-1 and weight-2 feature subsets and under two feature‑scaling strategies. Local eigentasks consistently improve performance relative to Pauli features; the gain is most pronounced when NSR‑aware scaling or NSR‑based selection is applied. We also report classical baselines trained directly on the raw Landsat features (horizontal reference lines). Baseline configurations follow standard \texttt{scikit‑learn} settings without tuning, except for $k$‑nearest neighbours ($k=7$) and MLP (\texttt{max\_iter} increased to ensure convergence); see Appendix~\ref{app:results/classical_baselines} for details.

\subsection{Sensitivity to the NSR cutoff}
\label{sec:results/landsat_cutoff}

\begin{figure}[ht]
    \centering
    \includegraphics[width=\linewidth]{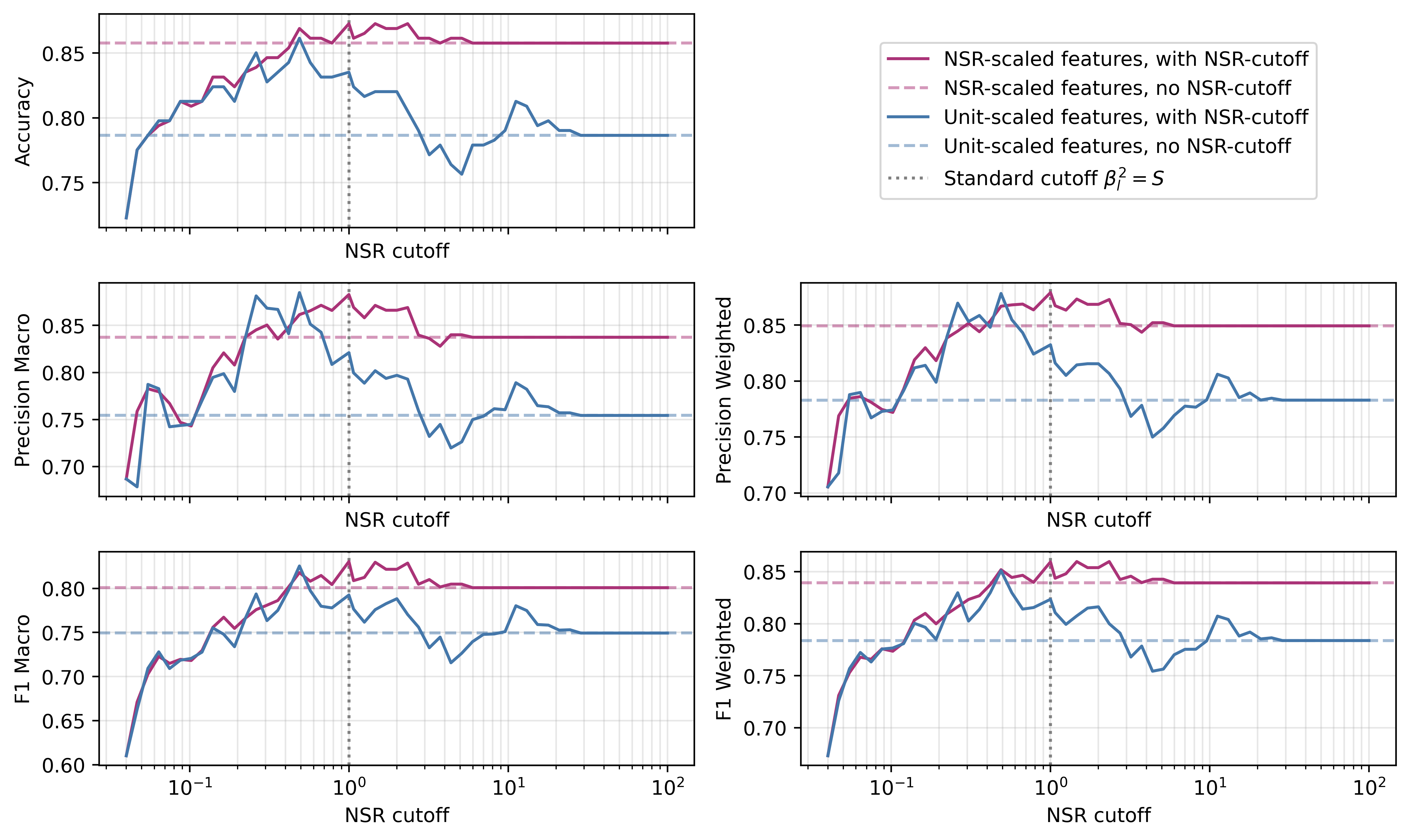}
    \caption{Sensitivity of Landsat classification performance to the NSR cutoff for 2‑local eigentasks (124‑qubit experiment). Metrics are plotted versus the threshold $\lambda$ in the criterion $\beta_l^2/S \le \lambda$. The vertical dashed line marks the theoretically motivated value $\lambda=1$ (signal equals noise). Horizontal dashed lines indicate the ``no‑cut'' baseline ($\lambda \to \infty$), where all eigentasks are retained.}
    \label{fig:eigentask124_vs_cutoff}
\end{figure}

We study how classification performance depends on the Noise-to-Signal Ratio (NSR) threshold used to discard unresolved eigentasks in the 2‑local readout. Let $S$ denote the shot budget and $\beta_l^2$ the NSR of the $l$-th eigentask. We define a cutoff parameter $\lambda > 0$ and retain eigentask $l$ if and only if $\beta_l^2/S < \lambda$.
In this parametrization, $\lambda=1$ corresponds to the theoretically motivated boundary where noise equals signal~\cite{hu_tackling_2023}. Figure~\ref{fig:eigentask124_vs_cutoff} reports standard classification metrics as a function of~$\lambda$.

Two consistent trends emerge. First, NSR‑aware scaled eigentasks are robust across a broad range of $\lambda$ and typically achieve their best performance when $\lambda$ is close to~$1$. Second, unit‑variance‑normalized features are more sensitive to the cutoff, reflecting the fact that normalization does not account for shot‑noise structure. As reference points, the rightmost markers correspond to retaining all eigentasks ($\lambda \to \infty$), while very small $\lambda$ removes almost all directions, leading to degraded performance.

\begin{figure}[ht]
    \centering
    \includegraphics[width=\linewidth]{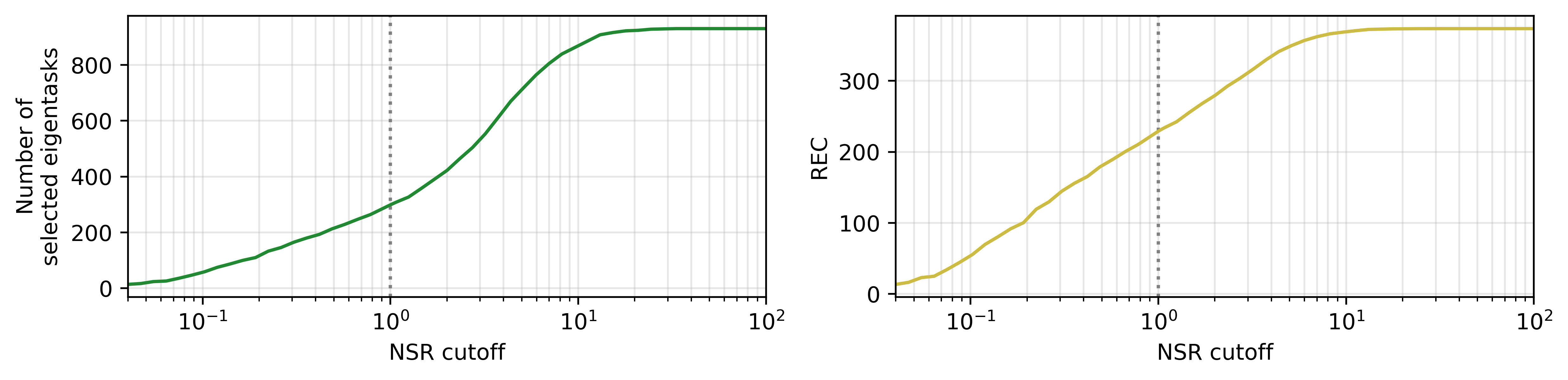}
    \caption{Effect of the NSR cutoff on readout size and resolvable content for 2‑local eigentasks (124‑qubit experiment). \textbf{Left:} number of retained eigentasks as a function of the threshold $\lambda$ in $\beta_l^2/S \le \lambda$. \textbf{Right:} resolvable expressive capacity (REC) $C_\lambda = \sum_{l:\,\beta_l^2/S < \lambda}\! 1/(1+\beta_l^2/S)$ compared to the total $C_T$. The vertical dashed lines highlight $\lambda=1$.}
    \label{fig:number_of_eigentasks_vs_cutoff}
\end{figure}

To quantify how the cutoff changes the readout dimension and the resolvable content, Fig.~\ref{fig:number_of_eigentasks_vs_cutoff} (left) shows the number of retained eigentasks as a function of~$\lambda$, and Fig.~\ref{fig:number_of_eigentasks_vs_cutoff} (right) reports the associated resolvable expressive capacity (REC)~\cite{hu_tackling_2023},
\begin{equation}
    C_{\text{total}} \;=\; \sum_l \frac{1}{1+\beta_l^2/S},
    \qquad
    C_\lambda \;=\; \sum_{l:\,\beta_l^2/S < \lambda} \frac{1}{1+\beta_l^2/S}.
\end{equation}
At the principled threshold $\lambda=1$, the 124‑qubit, 2‑local analysis retains 299 eigentasks out of a total of 930, striking a favorable balance between stability (shot‑noise robustness) and expressivity (feature richness).

\subsection{Classical baselines}
\label{app:results/classical_baselines}

For reference, we report Landsat results for a set of standard supervised‑learning models implemented in \texttt{scikit-learn}. For all models, the evaluation pipeline begins by normalizing the input features (centering and unit‑variance scaling). All models were used with their default hyperparameters, with two exceptions. First, for the $k$‑nearest neighbours classifier we set $k$ equal to the number of classes ($k=7$). Second, for the multilayer perceptron classifier we increased the maximum number of training iterations to ensure convergence. The corresponding configurations are reported in Table~\ref{tab:classical_baselines}.

\begin{table}[h]
\centering
\setlength{\tabcolsep}{5pt}
\begin{tabular}{lll}
\hline
\multirow{2}{*}{\textbf{Model}} & \multicolumn{2}{l}{\textbf{Hyperparameters}} \\
 & \textbf{Non-default} & \textbf{Default essential} \\
\hline

\verb|RidgeClassifierCV| &
/ &
\verb|fit_intercept=True| \\[3pt]

\verb|KNeighborsClassifier| &
\verb|n_neighbors=7| &
\verb|metric=`minkowski'|, \verb|p=2|, \verb|weights=`uniform'| \\[3pt]

\verb|SVC| &
/ &
\verb|kernel=`rbf'|, \verb|C=1|, \verb|gamma=`scale'| \\[3pt]

\verb|DecisionTreeClassifier| &
/ &
\verb|criterion=`gini'|, \verb|max_depth=None| \\[3pt]

\verb|RandomForestClassifier| &
/ &
\verb|n_estimators=100|, \verb|max_features=`sqrt'|, \verb|bootstrap=True| \\[3pt]

\verb|MLPClassifier| &
\verb|max_iter=1000| &
\verb|hidden_layer_sizes=(100,)|, \verb|activation=`relu'|, \verb|solver=`adam'| \\[3pt]

\verb|AdaBoostClassifier| &
/ &
\verb|n_estimators=50|, \verb|learning_rate=1|, \verb|algorithm=`SAMME.R'| \\[3pt]

\verb|GaussianNB| &
/ &
\verb|var_smoothing=1e-9| \\[3pt]

\verb|QuadraticDiscriminantAnalysis| &
/ &
\verb|reg_param=0|, \verb|store_covariance=False| \\
\hline
\end{tabular}
\caption{Classical baselines used for comparison.  
All hyperparameters are shown in \texttt{sklearn} code notation. The middle column lists deviations from defaults; the right column lists default settings relevant for reproducibility. Models were taken from \texttt{scikit-learn}~1.7.1.}
\label{tab:classical_baselines}
\end{table}

\end{document}